\documentclass[english,aps,prc,nofootinbib,superscriptaddress,twocolumn,letter,floatfix]{revtex4}
\usepackage[latin9]{inputenc}
\usepackage{hyperref}
\hypersetup{
  colorlinks=true,        
  linkcolor=blue,         
  citecolor=cyan,         
}
\usepackage{breakurl}
\usepackage{graphicx}
\usepackage{epsf}
\usepackage{epsfig}
\usepackage{amssymb,amsmath}
\usepackage[usenames]{color}
\usepackage{amssymb}
\usepackage{times}
\usepackage{comment}

\newcommand\snn{\sqrt{s_\text{NN}}}

\newcommand\pt{p_\text{T}}

\allowdisplaybreaks



\begin{document}

\preprint{This line only printed with preprint option}


\title{Left-right splitting of elliptic flow in heavy ion collisions: TRENTo-3D initialization and CLVisc hydrodynamic simulations}

\author{Ze-Fang Jiang}
\email{jiangzf@mails.ccnu.edu.cn}
\affiliation{Department of Physics and Electronic-Information Engineering, Hubei Engineering University, Xiaogan, Hubei, 432000, China}
\affiliation{Institute of Particle Physics and Key Laboratory of Quark and Lepton Physics (MOE), Central China Normal University, Wuhan, Hubei, 430079, China}

\author{Xiang Fan}
\email{xfan@mails.ccnu.edu.cn}
\affiliation{Institute of Particle Physics and Key Laboratory of Quark and Lepton Physics (MOE), Central China Normal University, Wuhan, Hubei, 430079, China}

\author{Duan She}
\email{sheduan@hnas.ac.cn}
\affiliation{Institute of Physics, Henan Academy of Sciences, Zhengzhou 450046, China}

\author{Shasha Ye}
\affiliation{Department of Physics and Electronic-Information Engineering, Hubei Engineering University, Xiaogan, Hubei, 432000, China}

\author{Ben-Wei Zhang}
\email{bwzhang@mail.ccnu.edu.cn}
\affiliation{Institute of Particle Physics and Key Laboratory of Quark and Lepton Physics (MOE), Central China Normal University, Wuhan, Hubei, 430079, China}

\begin{abstract}
Using the TRENTo-3D initial condition model coupled with (3+1)-dimensional CLVisc hydrodynamic simulations, we systematically investigate the left-right splitting of elliptic flow ($\Delta v_{2}$) for soft particles in relativistic heavy-ion collisions. Our study reveals that the final distribution characteristics of $\Delta v_{2}$ are primarily depend on the odd flow harmonics and $v_{2}$ itself. We find that the parton transverse momentum scale $k_\text{T}$ not only determines the geometric tilt of the QGP fireball but also significantly affects the rapidity dependence of both $v_1$ and $\Delta v_{2}$, providing new insights into the splitting mechanism of $\Delta v_{2}$. 
Furthermore, our results demonstrate that $\Delta v_{2} (p_\text{T})$ exhibits significant sensitivity to influences such as the  sub-nucleonic degrees of freedom (or `hotspots'), transverse momentum scale, and fragmentation region profile. By analyzing the $\Delta v_{2}$ and $\Delta v_{2}/v_{2}$ ratio, our findings provide new constraints on the uncertainties of the QGP initial state and provide additional constraints for refining model parameters.
\end{abstract}
\maketitle
\date{\today}

\section{Introduction}
\label{emsection1}
Hot and dense quantum chromodynamics (QCD) matter, known as the quark-gluon plasma (QGP), has been extensively studied in high-energy collisions at the Relativistic Heavy-Ion Collider (RHIC) and the CERN Large Hadron Collider (LHC)~\cite{Heinz:2013th,Busza:2018rrf}. Physicists found that at sufficiently high energies, the QGP exhibits nearly perfect fluid behavior~\cite{Song:2007fn,Petersen:2010cw,STAR:2003wqp,STAR:2014clz,STAR:2013qio}. In non-central heavy-ion collisions, the huge initial angular momentum renders the QGP the most vortical fluid ever observed~\cite{Liang:2004ph,STAR:2017ckg,STAR:2019erd}. Furthermore, an extremely strong magnetic field is generated due to the relativistic motion of the colliding nucleons beams~\cite{Fukushima:2008xe,Deng:2012pc,Deng:2017ljz,Zhao:2022dac}. These unique phenomena provide new insights into the properties of QCD matter.

Anisotropic flows, or collective flows, in the final-state particle momentum distribution are among the most fascinating phenomena in heavy-ion collisions, offering unique insights into the dynamics and properties of this strongly-coupled matter~\cite{STAR:2003wqp,ALICE:2010suc,ATLAS:2014txd,CMS:2012zex}. These flows include the directed flow ($v_{1}$), elliptic flow ($v_{2}$), triangular flow ($v_{3}$), and higher-order harmonics~\cite{Voloshin:2008dg,Bilandzic:2010jr}. Significant anisotropic flows have been observed not only in non-central heavy-ion collisions but also in small systems such as $p$+Au, $d$+Au, He+Au at RHIC~\cite{PHENIX:2018lia,STAR:2022pfn}, and $p$+Pb collisions at the LHC~\cite{CMS:2013jlh,Zhao:2020wcd,Tang:2023wcd}. Hydrodynamic models reveal that these flows are highly sensitive to the initial conditions of the QGP evolution~\cite{Ivanov:2005yw,Bozek:2011ua,Shen:2017bsr,Shen:2020jwv,Alzhrani:2022dpi,Jiang:2020big,Jiang:2021foj,Jia:2022ozr,Song:2017wtw}. Important insights have been gained into the impact of initial geometry, overlap fluctuations, and spatial anisotropies on the final-state anisotropic flow, driven by pressure gradients in hydrodynamic models~\cite{Pang:2018zzo,Wu:2021fjf,Shen:2020mgh} or particle interactions in AMPT transport models~\cite{Lin:2004en,Lin:2021mdn}.

Recently, the splitting of elliptic flow ($\Delta v_{2}$) on opposite sides of the impact parameter axis has emerged as a promising observable for constraining the initial state of the QGP~\cite{Chen:2021wiv,Zhang:2021cjt,Parida:2022lmt}. Studies within both transport and hydrodynamic frameworks suggest that $\Delta v_{2}$ is closely related to the non-zero angular momentum or the longitudinal tilted geometry of the initial QGP fireball, with the directed flow ($v_{1}$) being the primary contributor~\cite{Zhang:2021cjt,Parida:2022lmt}. In this study, we revisit $\Delta v_{2}(\eta)$ and $v_{2}(\pt)$ in non-central Au+Au collisions at $\sqrt{s_{NN}} = 200$ GeV. Using the state-of-the-art three-dimensional initial condition model TRENTo-3D~\cite{Soeder:2023vdn} to generate the initial energy density and coupling it with the (3+1)-dimensional CCNU-LBNL-Viscous hydrodynamic (CLVisc) model~\cite{Pang:2018zzo,Wu:2021fjf}, we explore the splitting of elliptic flow. Our results reveal that $\Delta v_{2}$ is primarily driven by odd flow harmonics, with $v_{1}$ being the dominant contributor to $\Delta v_{2}(\eta)$. However, unlike previous studies, we find that triangular flow ($v_{3}$) plays a significant role in $\Delta v_{2}(\pt)$, particularly in the pseudo-rapidity region $|\eta| < 1.3$.

To further investigate the sensitivity of $\Delta v_{2}$ to the initial state, we systematically examine the influence of sub-nucleonic degrees of freedom ($n_c$, or ``hotspots''), transverse momentum scale ($k_\text{T}$), and fragmentation profile parameters ($\alpha$ and $\beta$). We find that $\Delta v_{2}(\pt)$ is sensitive to all three effects in the TRENTo-3D initialization, while $\Delta v_{2}(\eta)$ exhibits strong sensitivity to $k_\text{T}$, which generates a tilted geometry of the QGP fireball, but minimal sensitivity to $n_c$, $\alpha$, and $\beta$. These findings suggest that $\Delta v_{2}$ could be a valuable observable for probing the initial state of the QGP fireball and refining parameters in the TRENTo-3D model. Additionally, we present predictions for $\Delta v_{2}/v_{2}$ ratio as a function of $\eta$ and $\pt$. We find that $\Delta v_{2}$ can complement $v_{1}$ and $v_{3}$ in constraining the three-dimensional properties of the QGP fireball created in heavy-ion collisions.

The article is organized as follows. In Sec.~\ref{v1section2}, we present the theoretical framework for investigating the splitting of elliptic flow, including the TRENTo-3D initialization with a tilted medium geometry, the (3+1)-dimensional hydrodynamic model for QGP evolution, and the mathematical formulation of $\Delta v_{2}$. In Sec.~\ref{v1section3}, we present and analyze the numerical results for $\Delta v_{2}$. A brief summary is provided in Sec.~\ref{v1section4}.

\section{Theory framework}
\label{v1section2}

\subsection{TRENTo-3D initialization} \label{sec:trento3d_detail}

We study Au+Au collisions at $\snn=$ 200 GeV. The initial condition for the hydrodynamic evolution of the QGP medium is obtained from the TRENTo-3D model~\cite{Soeder:2023vdn}.
Building on the TRENTo's participant thickness formalism, the three dimensional energy density profile at the initial proper time is 

\begin{equation}
\begin{aligned}
\varepsilon_{\text{IC}}(\vec{x}_\perp,\eta_s) &= \underbrace{N_{\text{fb}}\sqrt{T_A(\vec{x}_\perp)T_B(\vec{x}_\perp)}f_{\text{fb}}(\eta_s )}_{\text{Central fireball}} \\ 
&+ \sum_{X=A,B}\underbrace{\frac{k_\text{T}}{2N_{\text{frag}}}F_X(\vec{x}_\perp)f_{\text{frag}}^X(\eta_s)}_{\text{Fragmentation regions}}.    
\end{aligned}
\label{eq:tot_energy}
\end{equation}

The nuclear thickness functions $T_X$ incorporate sub-nucleonic structure through $n_c$ constituent partons with Gaussian smearing: 
\begin{equation}
T_X(\vec{x}_\perp) = \sum_{p\in X}\frac{1}{n_c}\sum_{c\in p}\gamma_c\frac{\exp\left(-\frac{|\vec{x}_\perp - \vec{x}_p - \vec{s}_c|^2}{2v^2}\right)}{2\pi v^2},
\end{equation}
where constituent's transverse positions $\vec{s}_c$ follow $\mathcal{N}(0,\sqrt{w^2 - v^2})$ with $v < w$, and fraction factor $\gamma_c$ obey the beta distribution:
\begin{equation}
\gamma_{c}\sim \text{Beta} \left\{ \langle\gamma_{c}\rangle \frac{1-k}{k}, (1-\langle\gamma_{c}\rangle) \frac{1-k}{k}\right\}.
\end{equation}

Central to TRENTo-3D's predictive power is its energy-momentum self-consistent deposition scheme. The fireball component's rapidity profile combines a deformed Gaussian with energy-dependent tapering:
\begin{equation}
f_{\text{fb}}(\eta_s) = \exp\left(-\frac{|\eta_s - \eta_{\text{cm}}|^f}{2\Delta\eta^2}\right)\left[1-\left(\frac{\eta_s - \eta_{\text{cm}}}{\eta_{\text{max}}}\right)^4\right]^4,
\end{equation}
where $\Delta\eta = \eta_{\text{max}} - \nu$ controls the Gaussian width and $\eta_{\text{cm}}$ ensures local momentum conservation:
\begin{equation}
\eta_{\text{cm}}(\vec{x}_\perp) = \text{arctanh}\left[\sqrt{1-\frac{4m_p^2}{s_{\text{NN}}}}\frac{T_A(\vec{x}_\perp) - T_B(\vec{x}_\perp)}{T_A(\vec{x}_\perp) + T_B(\vec{x}_\perp)}\right],
\end{equation}

The fragmentation regions in Eq.~(\ref{eq:tot_energy}) are described using modified parton distribution functions scaled by the transverse momentum scale $k_\text{T}$:
\begin{equation}
f_{\text{frag}}^X(\eta_s) = (-\ln x)^\alpha x^{\beta+1}\exp\left(-\frac{2k_\text{T}}{x\sqrt{s_{\text{NN}}}}\right),\ \ x = e^{-(\eta_{\text{max}} \mp \eta_s)},
\end{equation}
where the $-$/$+$ signs correspond to the forward/backward regions, respectively. The thickness functions for the $X$-going fragments are given by:
\begin{equation}
F_X(\vec{x}_\perp) = \sum_{p\in X}\frac{1}{n_c}\sum_{c\in p}(1-\gamma_c)\frac{\exp\left(-\frac{|\vec{x}_\perp - \vec{x}_p - \vec{s}_c|^2}{2v^2}\right)}{2\pi v^2}.
\end{equation}
The dynamic rapidity window $\eta_{\text{max}}$ evolves with collision energy as:
\begin{equation}
\eta_{\text{max}} = \text{arccosh}\left(\frac{\sqrt{s_\text{NN}}}{2k_\text{T}}\right) \approx \ln\left(\frac{\sqrt{s_\text{NN}}}{k_\text{T}}\right) - \ln 2 + \mathcal{O}\left(\frac{k_\text{T}^2}{s_\text{NN}}\right).
\end{equation}

The longitudinal structure of the QGP fireball in the TRENTo-3D initialization is characterized by both central and fragmentation regions, governed by the following key parameters~\cite{Soeder:2023vdn}:

\begin{table}[ht]
\caption{Parameter values used in this study~\cite{Soeder:2023vdn}.}
\label{tab:params3d}
\centering\footnotesize
\begin{tabular}{@{}llp{5.8cm}@{}}
\toprule
Parameter & Value & Influence \\
\hline
$n_c$ & 16.4 & Tunes initial subnucleonic degrees of freedom \\
$w$ [fm] & 1.3 & Sets initial source size $\propto \sqrt{\langle r^2 \rangle}$ \\
$\chi = v/(w n_c^{1/4})$ & 0.5 & Tunes subnucleon correlation length \\
$f$ & 1.0 & Controls the central fireball profile \\
$k_\text{T}$ [GeV] & 0.33 & Determines $\eta_{\text{max}}$ and tilted geometry \\
$\alpha$ & 4.6 & Controls the shape of the fragmentation profile \\
$\beta$ & 0.19 & Controls the shape of the fragmentation profile \\
\toprule
\end{tabular}
\end{table}

In this work, we focus on the influence of subnucleonic degrees of freedom ($n_c$), the transverse momentum scale ($k_\text{T}$), and the fragmentation profile parameters ($\alpha$ and $\beta$) on the elliptic flow splitting ($\Delta v_{2}$) of charged particles. The results and detailed analysis are presented in Section~\ref{v1section3}. The energy deposition in the model dynamically adapts according to:
\begin{equation}
\langle\gamma\rangle\sqrt{s_{\text{NN}}} = 2N_{\text{fb}} \int_{-\eta_{\text{max}}}^{\eta_{\text{max}}} \cosh\eta_s f_{\text{fb}}(\eta_s) d\eta_s,
\end{equation}
which captures the transition from strong net-baryon accumulation at RHIC energies to midrapidity-dominated deposition at LHC energies. The TRENTo-3D initialization parameters used in this study are based on Ref.~\cite{Soeder:2023vdn}, which have been validated through Bayesian calibration to $dN_{\text{ch}}/d\eta$ distributions across the energy range of 200 GeV--5.02 TeV. Those parameters successfully reproduces the $dN_{\text{ch}}/d\eta$ distributions in both p+A and A+A collisions.

\begin{figure*}[tbp!]
\begin{center}
\includegraphics[width=0.9\linewidth]{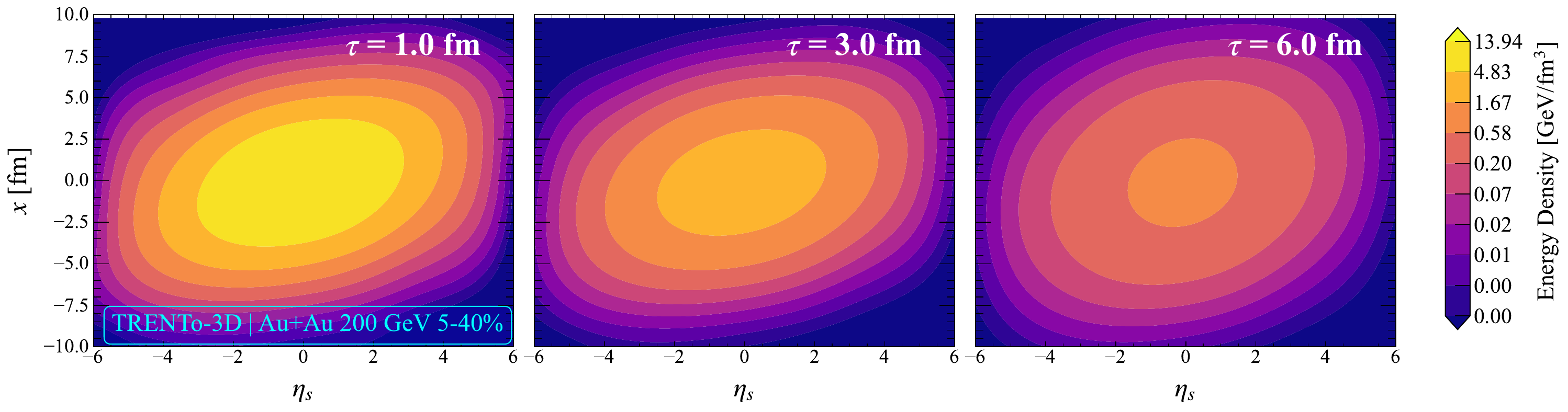}~
\end{center}
\caption{(Color online) 
The energy density evolution in the reaction plane ($\eta_\text{s}$-$x$ plane) for 5-40\% Au+Au collisions at $\sqrt{s_\text{NN}}=200$~GeV. The initial energy profile from the TRENTo-3D initialization averaged 2000 events (one-shot).}
\label{f:ed}
\end{figure*}

\subsection{CLVisc hydrodynamic simulations}

Starting with the TRENTo-3D initial condition constructed in the previous subsection, we utilize a (3+1)-D viscous hydrodynamic model CLVisc to describe the evolution of the QGP medium and further particlization. 
The CLVisc model is described in full detail in Refs.~\cite{Pang:2018zzo,Wu:2021fjf}. We repeat here only its main features. The CLVisc code solves the local energy-momentum conservation equations 
\begin{align}
\nabla_{\mu} T^{\mu\nu}&=0 \, ,\\
\nabla_{\mu} J^{\mu}&=0  \, ,
\end{align}
where the energy-momentum tensor $T^{\mu\nu}$ and the net baryon current $J^{\mu}$ are defined as
\begin{align}
T^{\mu\nu} &= \varepsilon U^{\mu}U^{\nu} - P\Delta^{\mu\nu} + \pi^{\mu\nu}\,, \\	
J^{\mu} &= nU^{\mu}+V^{\mu}\,,
\end{align}
with $\varepsilon$, $P$, $n$, $u^{\mu}$, $\pi^{\mu\nu}$, $V^{\mu}$ being the local energy density, pressure, net baryon density (which assumed $n=0$ in this study), flow velocity field, shear stress tensor and baryon diffusion current respectively. 
The projection tensor is given by $\Delta^{\mu\nu} = g^{\mu\nu}-u^{\mu}u^{\nu}$ with the metric tensor $g^{\mu\nu} = \text{diag} (1,-1,-1,-1)$. The dissipative currents $\pi^{\mu\nu}$ and $V^{\mu}$ are given by the Israel-Stewart-like second order hydrodynamic expansion~\citep{Denicol:2018wdp} and the effects of the bulk viscosity is not included in the present study yet.

We implement the HotQCD2014 equation of state (EOS) to solve the hydrodynamic equations ~\cite{Monnai:2019hkn,Monnai:2021kgu}. The freeze-out process is implemented at a critical energy density $e_{\text{frz}}$= 0.4~GeV/fm$^3$, where we construct the freeze-out hypersurface and perform hadron sampling using the Cooper-Frye formalism~\cite{Pang:2018zzo,Wu:2021fjf}. 
The resonance decay contributions also included in the simulation. It is worth noting that the present framework does not account for hadronic rescattering effects during the later-stage evolution of the system.

Utilizing the aforementioned parameters, we present in Fig.~\ref{f:ed} the energy density distributions density in the $\eta_\text{s}$-$x$ plane for a single event  (somth, averaged over 2000 events) in 5-40\% centrality Au+Au collisions at $\snn=200$~GeV, evaluated at proper time $\tau=1,~3,~6$ fm (from left to right panel). The figure reveals that the QGP fireball manifests a distinct tilted and asymmetric geometry along the longitudinal direction during its evolution under the TRENTo-3D initialization + CLVisc hydrodynamic framework. This geometric distortion is anticipated to be more pronounced in peripheral collisions, owing to enhanced drag forces exerted by spectator nucleons on participants. Such energy density asymmetry plays a crucial role in generating rapidity-odd directed flow of soft hadrons\cite{Bozek:2022svy}, providing valuable insights into the initial state dynamics of heavy-ion collisions.

\subsection{Left-right splitting of the elliptic flow }

The normalized azimuthal distribution of hadrons in momentum space, characterized by a given transverse momentum $p_{T}$ and either rapidity $y$ or pseudorapidity $\eta$, can be expressed as~\cite{Voloshin:2008dg,Zhang:2021cjt}  

\begin{equation}
\begin{aligned}
\frac{dN}{d\phi} = \frac{1}{2\pi}(1 &+ 2\sum_{n=1}^{\infty}(v_n\cos(n(\phi-\psi_{\textrm{RP}})) \\
                    &+ s_{n} \sin(n(\phi-\psi_{\textrm{RP}})))),
\label{eq16}
\end{aligned}
\end{equation}
where $\psi_{\textrm{RP}}$ is the angle of the reaction plane in the laboratory frame, $v_{n}$ and $s_{n}$ represent the Fourier coefficients that quantify the azimuthal anisotropy of the particle distribution.

A novel approach has been recently proposed to quantify the right (shorted as ``$R$'')-left (shorted as ``$L$'') elliptic flow splitting $\Delta v_{2}$ in $v_{2}= \left<\cos(2(\phi-\psi_{\textrm{RP}})) \right>$~\cite{Zhang:2021cjt}, where measurements are performed across distinct regions of final-state hadron momentum space:
\begin{equation}
\begin{aligned}
\Delta v_{2}^{\mathrm{RP}} &= v_{2}^{R} - v_{2}^{L},
\end{aligned}
\end{equation}
where $v_{2}^{R} = \left<\cos(2(\phi^{R} - \psi_{\textrm{RP}})) \right>$ for $\phi^{R} \in ((\psi_{\textrm{RP}} - \pi/2), (\psi_{\textrm{RP}} + \pi/2) )$, and $v_{2}^{L} = \left<\cos(2(\phi^{L} - \psi_{\textrm{RP}})) \right>$ for $\phi^{L} \in ((\psi_{\textrm{RP}} + \pi/2), (\psi_{\textrm{RP}} + 3\pi/2) )$. Since the reaction plane $\psi_\textrm{RP}$ cannot be directly measured in the experiment, the second-order event plane $\psi_{2}$ and the first order spectator plane $\psi_{\textrm{SP}}$ have been suggested as effective proxies~\cite{STAR:2013ksd,ALICE:2012nhw}. For $\Delta v_{2}$ determination, $\psi_{\textrm{SP}}$ is particularly advantageous as it resolves the phase distinct between $\phi^{R}$ and $\phi^{L}$, unlike $\psi_{2}$, which treats $\psi_{2} = \pi$ and $\psi_{2} = 0$ as equivalent. In addition, the recent installation of the Event Plane Detector at large rapidities at RHIC provides an alternative method for exactly estimating $\psi_{\textrm{RP}}$~\cite{Adams:2019fpo}.

If we integrate over $\phi^{R} \in ((\psi_{\textrm{RP}} - \pi/2), (\psi_{\textrm{RP}} + \pi/2))$ for the $p_{x} > 0$ part or over $\phi^{L} \in ((\psi_{\textrm{RP}} + \pi/2), (\psi_{\textrm{RP}} + 3\pi/2) )$ for the $p_{x} < 0$ part, we obtain the following for $v_{2}$ (with terms up to $n$ = 5):
\begin{equation}
\begin{aligned}
  v_{2}^{R} &= \frac{\int_{\psi_{\mathrm{RP}}-\frac{\pi}{2}}^{\psi_{\mathrm{RP}}+\frac{\pi}{2}} \cos\left[2(\phi - \psi_\mathrm{RP})\right] \frac{dN}{d\phi} d\phi}{\int_{\psi_{\mathrm{RP}}-\frac{\pi}{2}}^{\psi_{\mathrm{RP}}+\frac{\pi}{2}} \frac{dN}{d\phi} d\phi} \\
&\approx \frac{v_{2} + \frac{4}{3\pi} v_{1} + \frac{12}{5\pi}v_{3} - \frac{20}{21\pi}v_{5} }{ 1 + \frac{4}{\pi} v_{1} - \frac{4}{3\pi}v_{3} + \frac{4}{5\pi}v_{5}} \\
&= v_{2} \left [\frac{1+ \frac{4 v_{1}}{3\pi v_{2}} + \frac{12 v_{3}}{5\pi v_{2}} - \frac{20 v_{5} }{21\pi v_{2}} }{1 + \frac{4}{\pi} v_{1} - \frac{4}{3\pi}v_{3} + \frac{4}{5\pi}v_{5}} \right],
\label{eq16}
\end{aligned}
\end{equation}

\begin{equation}
\begin{aligned}
v_{2}^{L} &= \frac{\int_{\psi_{\mathrm{RP}}+\frac{\pi}{2}}^{\psi_{\mathrm{RP}}+\frac{3\pi}{2}} \cos\left[2(\phi - \psi_\mathrm{RP})\right] \frac{dN}{d\phi} d\phi}{\int_{\psi_{\mathrm{RP}}+\frac{\pi}{2}}^{\psi_{\mathrm{RP}}+\frac{3\pi}{2}} \frac{dN}{d\phi} d\phi} \\
&\approx \frac{v_{2} - \frac{4}{3\pi} v_{1} - \frac{12}{5\pi}v_{3} + \frac{20}{21\pi}v_{5} }{ 1 - \frac{4}{\pi} v_{1} + \frac{4}{3\pi}v_{3} - \frac{4}{5\pi}v_{5}} \\
&=v_{2} \left[\frac{1- \frac{4 v_{1} }{3\pi v_{2}}  - \frac{12 v_{3} }{5\pi v_{2}} + \frac{20 v_{5} }{21\pi v_{2}} }{ 1 - \frac{4}{\pi} v_{1} + \frac{4}{3\pi}v_{3} - \frac{4}{5\pi}v_{5}} \right].
\label{eq17}
\end{aligned}
\end{equation}

The splitting of elliptic flow, $\Delta v_{2}^{\text{RP}}$, can be written as 
\begin{equation}
\begin{aligned}
\Delta v_{2}^{\mathrm{RP}} &= v_{2}^{R} - v_{2}^{L} \\
  &= \frac{\frac{8}{3\pi}\left[v_{1}(1-3v_{2}) + (\frac{9}{5}+v_{2})v_{3} - (\frac{5}{7}+\frac{3}{5}v_{2})v_{5}\right]}{( 1 - \frac{4}{\pi} v_{1} + \frac{4}{3\pi}v_{3} - \frac{4}{5\pi}v_{5})( 1 + \frac{4}{\pi} v_{1} - \frac{4}{3\pi}v_{3} + \frac{4}{5\pi}v_{5})} \\ 
  & = \frac{\frac{8}{3\pi}\left[v_{1}(1-3v_{2}) + (\frac{9}{5}+v_{2})v_{3} - (\frac{5}{7}+\frac{3}{5}v_{2})v_{5}\right]}{1-\frac{16v_{1}^{2}}{\pi^{2}} + \frac{32 v_{1} v_{3}}{3\pi^{2}} -\frac{16v_{3}^{2}}{9\pi^{2}} - \frac{32v_{1}v_{5}}{5\pi^{2}}+\frac{32 v_{3}v_{5}}{15 \pi^{2}} - \frac{16v_{5}^{2}}{25\pi^{2}}}.
\label{eq18}
\end{aligned}
\end{equation}
When considering only the first - and second - order contributions, i.e., by setting \( v_3 = v_5 = 0 \), one obtains:
\begin{equation}
\begin{aligned}
\Delta v_{2,\textrm{1st}}^{\text{RP}} 
   & = \frac{\frac{8}{3\pi}\left[v_{1}(1-3v_{2}) \right]}{1-\frac{16v_{1}^{2}}{\pi^{2}}} 
    \approx \frac{8}{3\pi}v_{1}(1-3v_{2}). 
\label{eq19}
\end{aligned}
\end{equation}

When considering contributions up to the third order, with \( v_5 = 0 \), one obtains
\begin{equation}
\begin{aligned}
\Delta v_{2,\textrm{3rd}}^{\text{RP}} 
   &= \frac{\frac{8}{3\pi}\left[v_{1}(1-3v_{2}) + (\frac{9}{5}+v_{2})v_{3} \right]}{1-\frac{16v_{1}^{2}}{\pi^{2}} + \frac{32 v_{1} v_{3}}{3\pi^{2}} -\frac{16v_{3}^{2}}{9\pi^{2}}}  \\
   & \approx \frac{8}{3\pi}\left[v_{1}(1-3v_{2}) + (\frac{9}{5}+v_{2})v_{3} \right]. 
\label{eq20}
\end{aligned}
\end{equation}

And up to the fifth order, one obtains:
\begin{equation}
\begin{aligned}
\Delta v_{2,\textrm{5th}}^{\text{RP}} 
   &= \frac{\frac{8}{3\pi}\left[v_{1}(1-3v_{2}) + (\frac{9}{5}+v_{2})v_{3} - (\frac{5}{7}+\frac{3}{5}v_{2})v_{5}\right]}{1-\frac{16v_{1}^{2}}{\pi^{2}} + \frac{32 v_{1} v_{3}}{3\pi^{2}} -\frac{16v_{3}^{2}}{9\pi^{2}} - \frac{32v_{1}v_{5}}{5\pi^{2}}+\frac{32 v_{3}v_{5}}{15 \pi^{2}} - \frac{16v_{5}^{2}}{25\pi^{2}}} \\
   & \approx \frac{8}{3\pi}\left[v_{1}(1-3v_{2}) + (\frac{9}{5}+v_{2})v_{3} - (\frac{5}{7}+\frac{3}{5}v_{2})v_{5}\right].
\label{eq21}   
\end{aligned}
\end{equation}

It is evident that Eq.(\ref{eq19}) is consistent with the expression in Ref.\cite{Zhang:2021cjt}, while Eq.(\ref{eq21}) aligns with that in Ref.\cite{Parida:2022lmt} upon neglecting the ``$v_{odd}\cdot v_{2}$'' terms.
One can also see that $\Delta v_{2}^{\mathrm{RP}}$ depends on the odd flow harmonics and $v_{2}$ itself. Given that the odd flow harmonics exhibit rapidity-odd components relative to the reaction plane, $\Delta v_{2}^{\mathrm{RP}}$ inherits this rapidity-odd behavior.

Prior to calculating the elliptic flow splitting $v^{\textrm{RP}}_{2}$, it is important to note that the $v_{3}$ and $v_{5}$ does not represent the usual flow coefficient $v_{3}$ and $v_{5}$~\cite{Zhang:2021cjt}, which axes fluctuate independently of the $x$-axis (or the reaction plane). Instead, $v_{3}$ and $v_{5}$ here is calculated relative to the reaction plane and represents another type of third and fifth flow coefficient ($v_{3}^{\text{RP}}$ and $v_{5}^{\text{RP}}$) at finite rapidity that correlates with the reaction plane.

\section{Numerical results}
\label{v1section3}

In this section, we present our calculation and prediction for $\Delta v_{2}$ with respect to the reaction plane, using the TRENTo-3D initial condition and CLVisc hydrodynamic simulation. These results are compared with pre-existing experimental data.
 
\begin{figure}[tbp!]
\begin{center}
\includegraphics[width=0.85\linewidth]{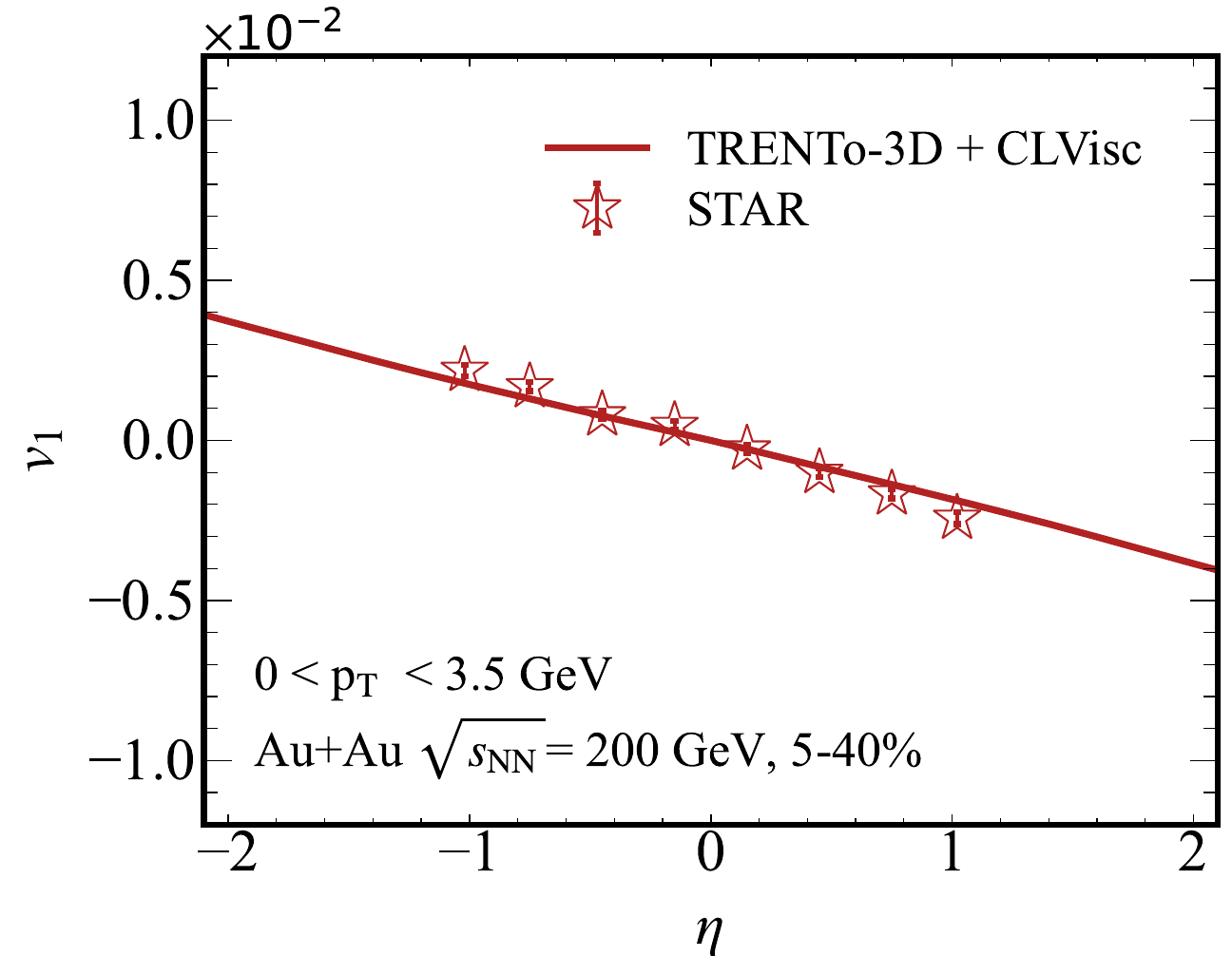} \\
\includegraphics[width=0.85\linewidth]{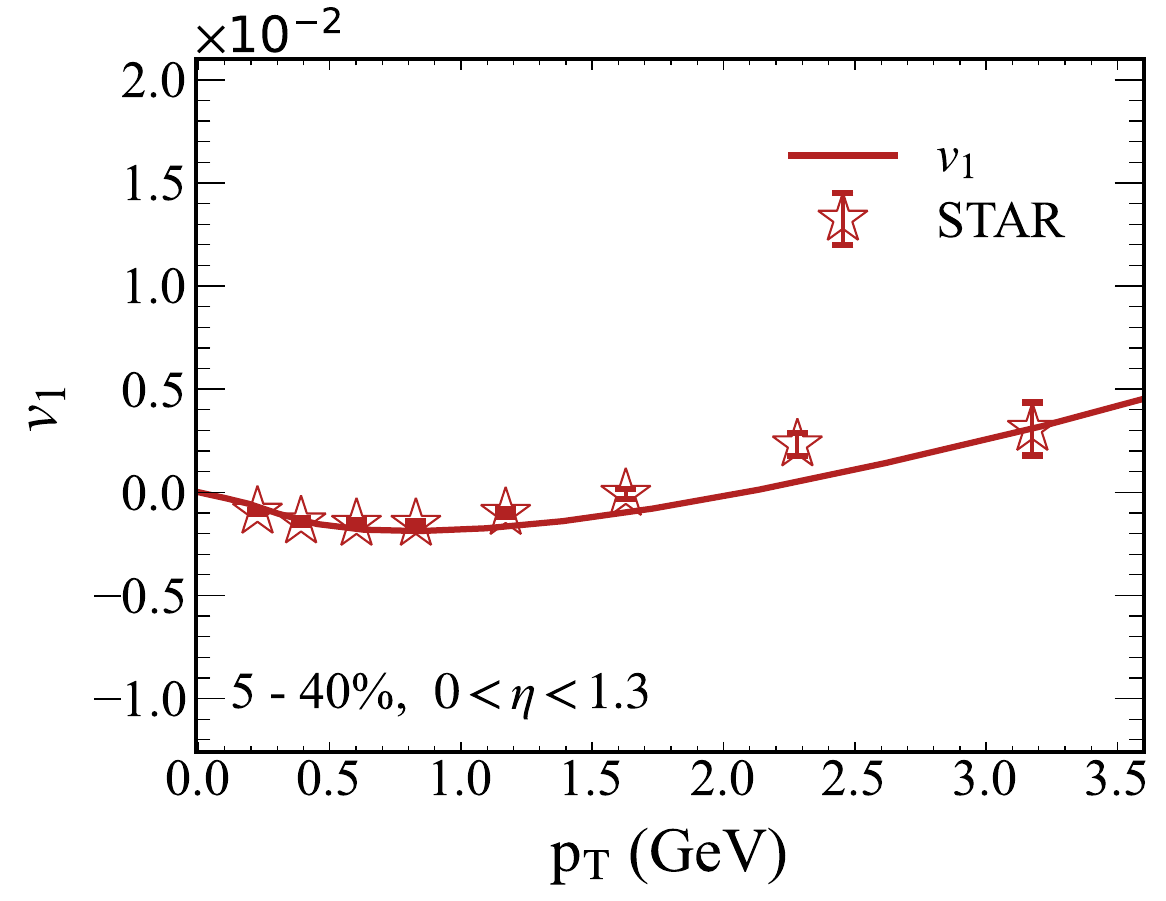}
\end{center}
\caption{(Color online) Upper panel: the rapidity dependence of the directed flow ($v_{1}$) of charged particle as obtained in the CLVisc hydrodynamic simulations with the TRENTo-3Dinitial condition (2000 events averaged) are compared to the experimental measurements. Lower panel: the transverse momentum dependence of the $v_{1}$ in 5-40\% centrality Au+Au collisions at 200 GeV. The experimental data are taken from the STAR collaboration~\cite{STAR:2014clz}.}
\label{f:v1}
\end{figure}

In Fig.~\ref{f:v1}, we present the directed flow ($v_1$) of charged particles as a function of pseudo-rapidity (upper panel) and transverse momentum (lower panel) in 5-40\% Au+Au collisions at $\snn$=200 GeV.
The calculations are carried out using the TRENTo-3D smooth initial condition (averaged 2000 events) and (3+1)-D CLVisc hydrodynamic simulations. 
We find that TRENTo-3D + CLVisc framework provide a reasonable description for the STAR experimental data. This suggests that the directed flow is influenced by the tilted geometry of the QGP fireball, aligning with previous research~\cite{Zhang:2021cjt,Parida:2022lmt}. Thus, we expect TRENTo-3D + CLVisc framework to provide correct prediction and deep understanding of $\Delta v_{2}$.

\begin{figure}[!t]
\includegraphics[width=0.85\linewidth]{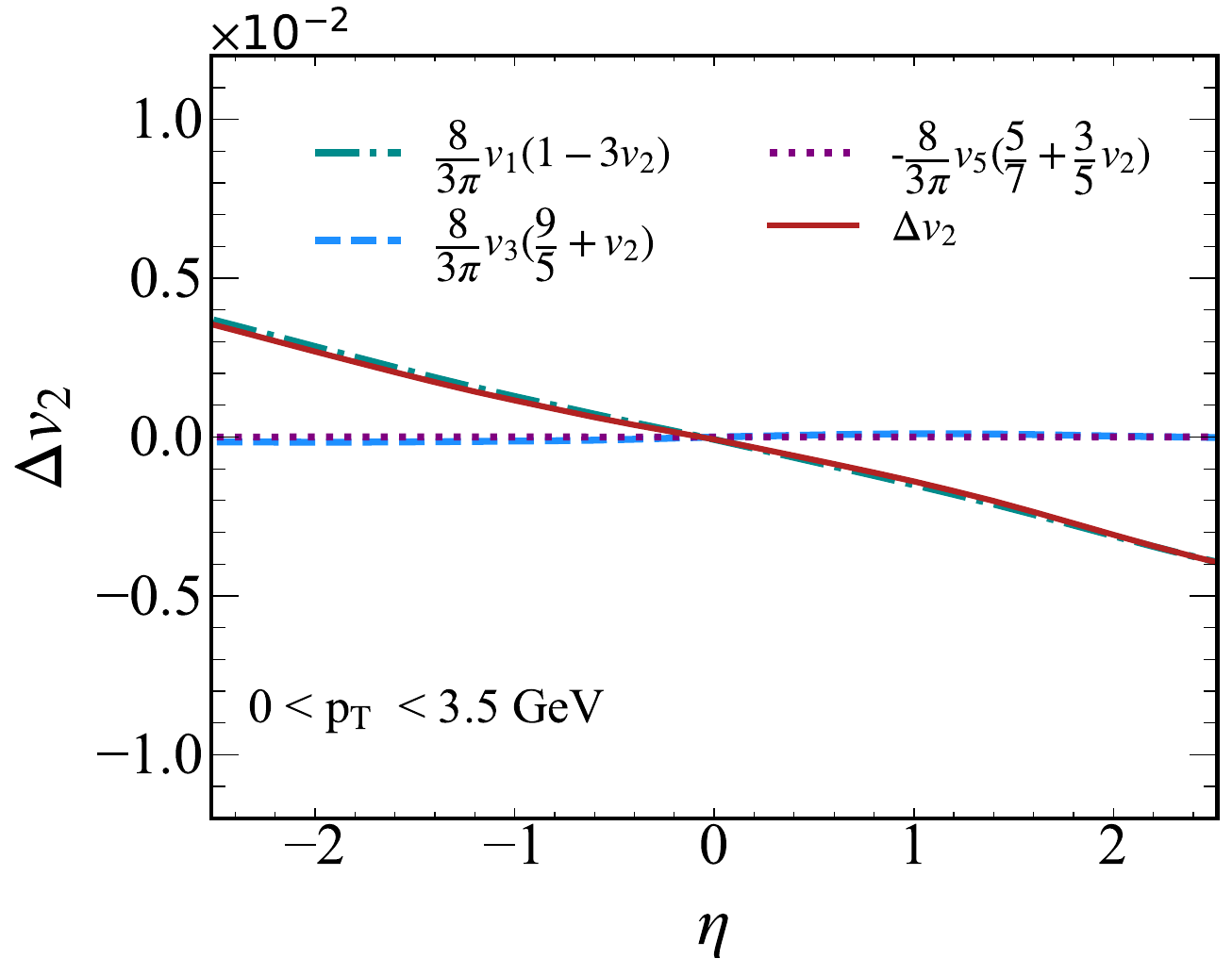}
\includegraphics[width=0.85\linewidth]{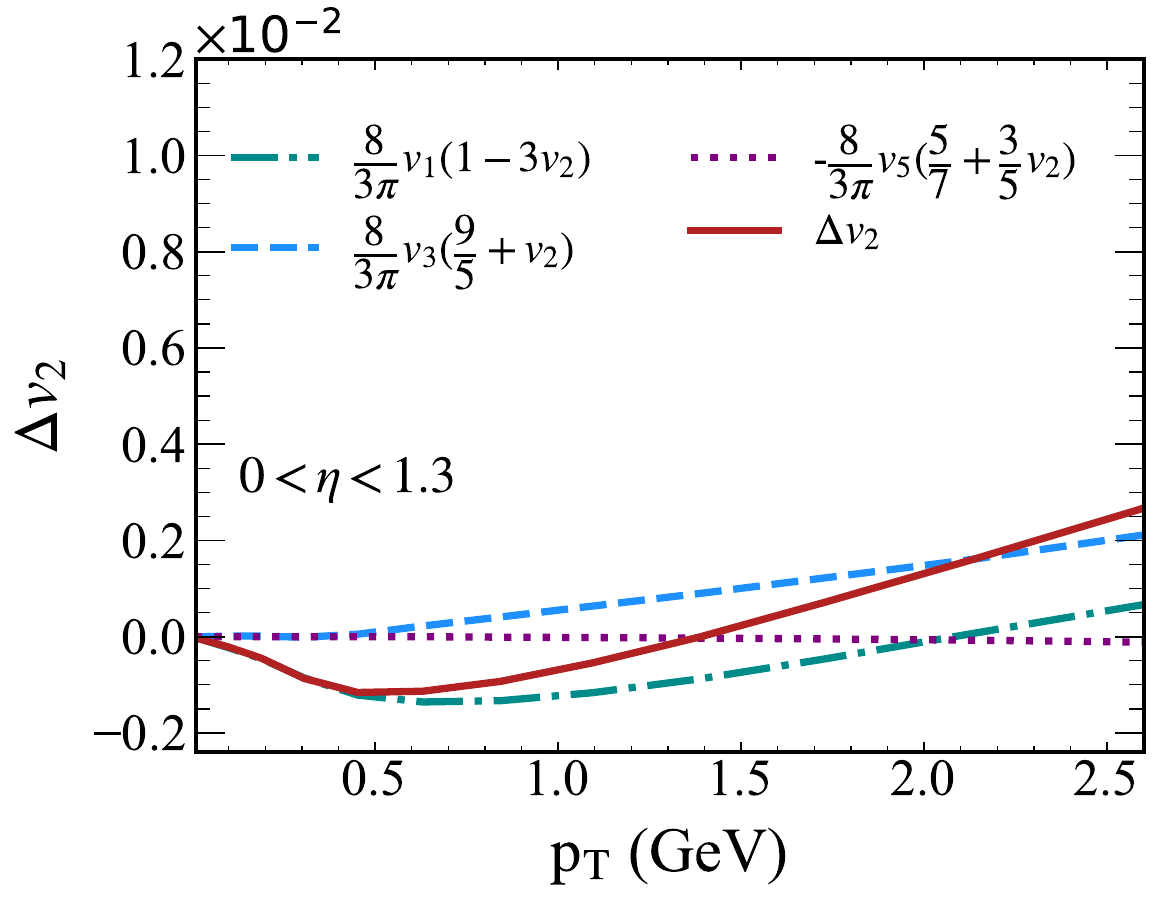}
\caption{(Color online) Upper panel:  Predicted $\Delta v_{2}$ as a function of $\eta$ for 5-40\% centrality Au+Au collisions at $\snn=200$~GeV.
Lower panel: Predicted $\Delta v_{2}$ as a function of $\pt$. Contributions to $\Delta v_{2}$ from odd flow harmonics are also displayed. 
}
\label{f:v1_cuau}
\end{figure}
  
In the upper panel of Fig.~\ref{f:v1_cuau}, we display the model prediction for elliptic flow splitting ($\Delta v_{2}$) of charged particles as a function of pseudorapidity ($\eta$) in mid-central (5-40\%) Au+Au collisions at $\snn=200$~GeV. We additionally plot odd harmonics with fixed coefficients as presented in Eq.~(\ref{eq21}), which constitute the dominant contribution to $\Delta v_{2}$. We find that the splitting primarily originates from the term $\frac{8v_{1}}{3\pi}(1-3v_{2})$, with $\Delta v_{2}$ showing equivalent sign dependence. The contributions to $\Delta v_{2}(\eta)$ from $v_{3}$ and $v_{5}$ harmonics, though sign-opposed to $v_{1}$, remain negligible over the full $\eta$ range.

In the lower panel of Fig.~\ref{f:v1_cuau}, we present the $\Delta v_{2}$ dependence on transverse momentum ($p_{T}$) at mid-rapidity ($|\eta|< 1.3$) in Au+Au collisions. It is interesting to see that $v_{3}$ constitutes the dominant contribution to $\Delta v_{2}(\pt)$, while contributions from $v_{5}$ harmonics remain negligible. A cancellation effect emerges at low $\pt$ ($\pt<1.2$ GeV) for $\Delta v_{2}(\pt)$, due to the negative $v_{1}$ and positive $v_{3}$ yield at low $\pt$. Then enhanced magnitude of $\Delta v_{2}(\pt)$ compared to prior studies~\cite{Zhang:2021cjt,Parida:2022lmt} arises from a different longitudinally tilted initial geometric configuration in TRENTo-3D initialization. 

\begin{figure}[!t]
\includegraphics[width=0.85\linewidth]{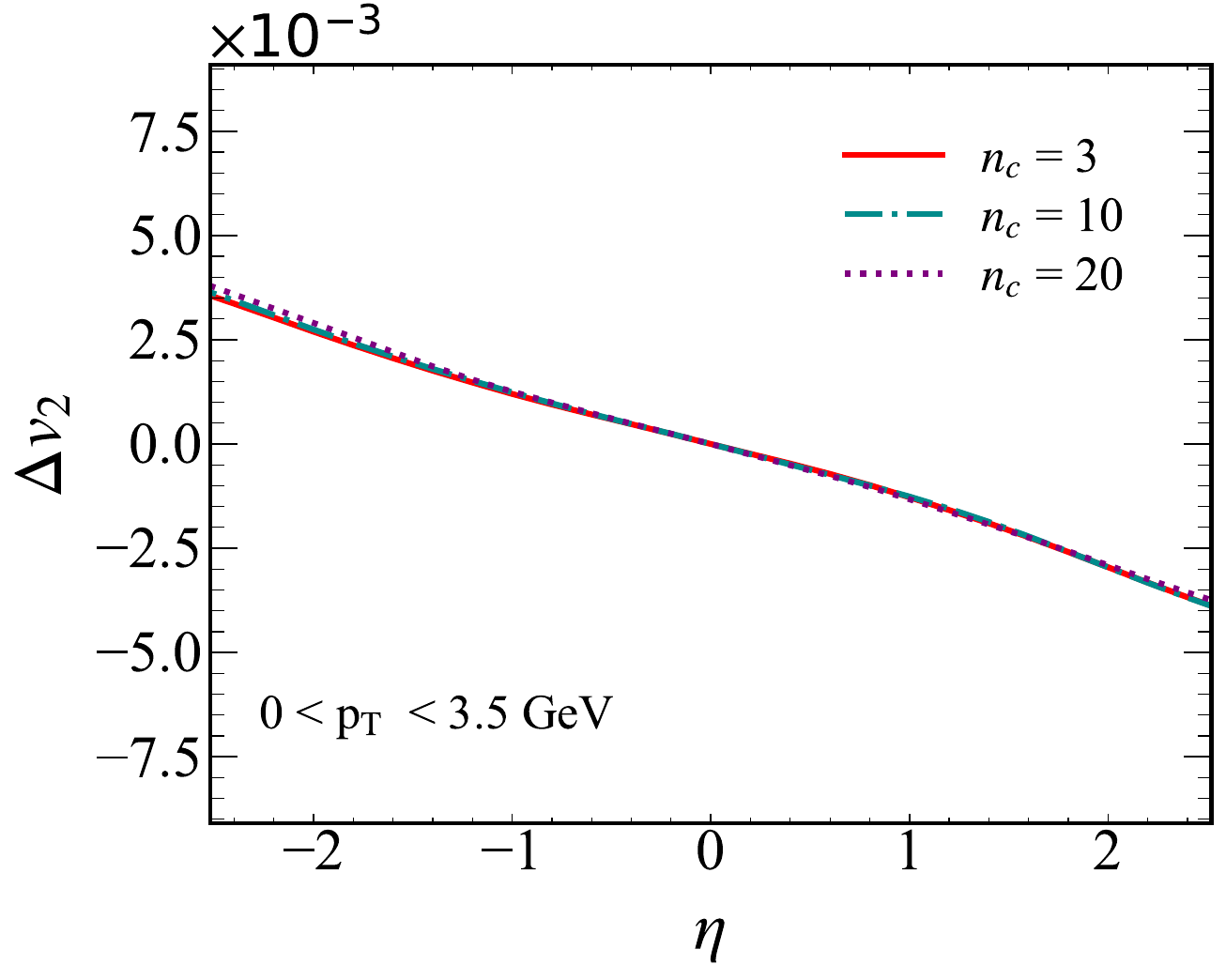}
\includegraphics[width=0.85\linewidth]{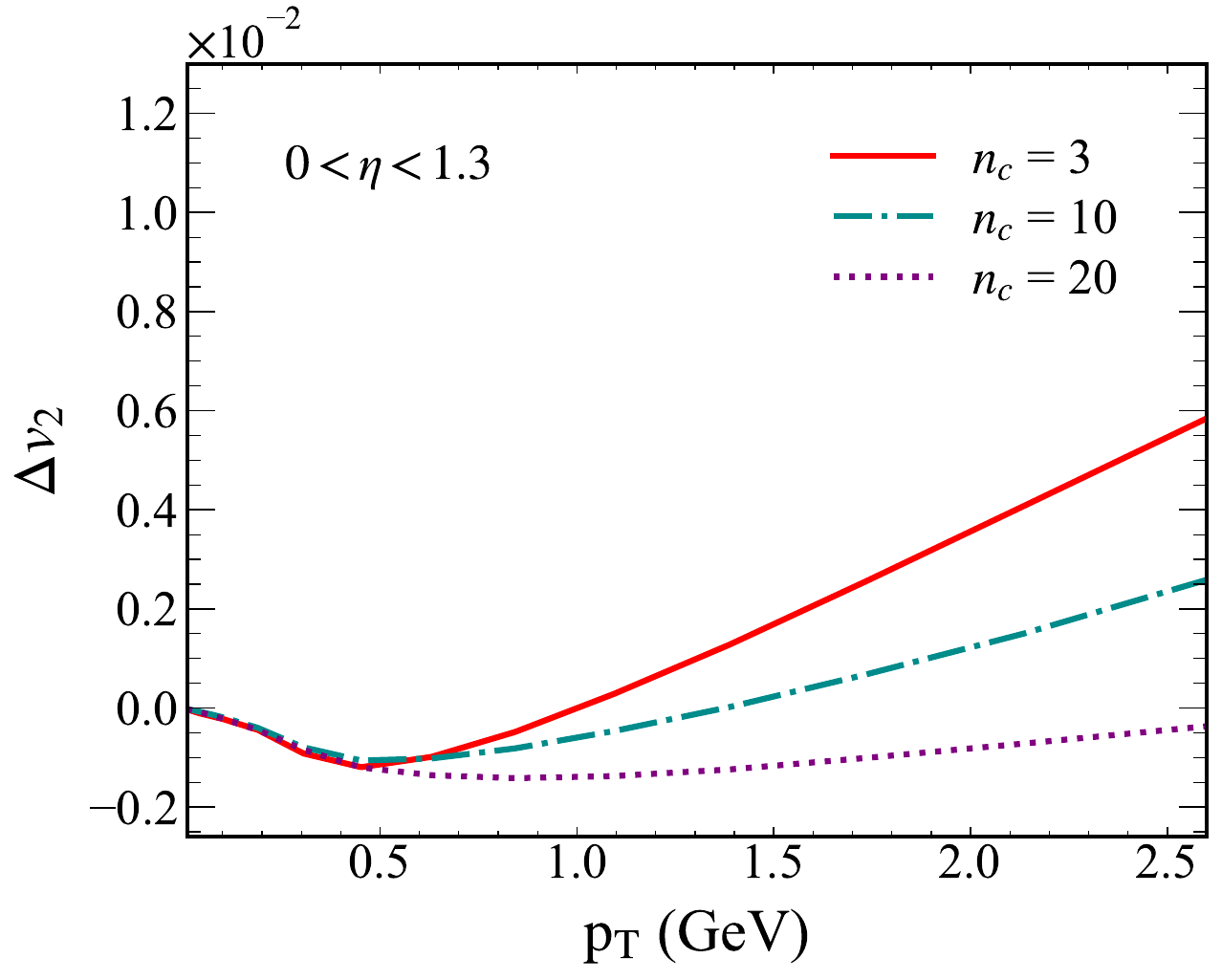}
\caption{(Color online) $\Delta v_{2}(\eta)$ (upper panel) and $\Delta v_{2}(\pt)$ (lower panel) distributions in Au+Au collisions at $\snn=200$~GeV (5-40\% centrality) for sub-nucleonic structure parameters $n_{c}=3,~10,~20$. }
\label{f:v1_nc}
\end{figure}

We now examine the influence of the number of constituents (or ``hotspots'') per nucleon ($n_{c}$), transverse momentum scale ($k_\text{T}$), and fragmentation profile parameters ($\alpha$ and $\beta$) on the elliptic flow splitting $\Delta v_{2}$.

We first investigate the impact of the number of constituents ($n_{c}$) on $\Delta v_{2}$. The upper panel of Fig.~\ref{f:v1_nc} displays the $\Delta v_{2}(\eta)$ distributions for Au+Au collisions at $\snn= 200$~GeV (5--40\% centrality, $0 < \pt < 3.5$~GeV) for different $n_c$ values. The smooth initial condition suppresses spatial fluctuations of sub-nucleonic structures through transverse-plane averaging, resulting in minimal impact on $\Delta v_{2}(\eta)$. In the lower panel, we present $\Delta v_{2}$ as a function of $\pt$. Enhanced ``hotspot'' fluctuations at larger $n_c$ correlate with decreased $\Delta v_2(\pt)$, suggesting a connection between $\Delta v_2$ enhancement and the growth of triangular flow ($v_3$) under stronger initial-state hotspot fluctuations, as indicated in previous studies~\cite{Zhu:2024tns,Garcia-Montero:2025bpn}.

\begin{figure}[!t]
\includegraphics[width=0.85\linewidth]{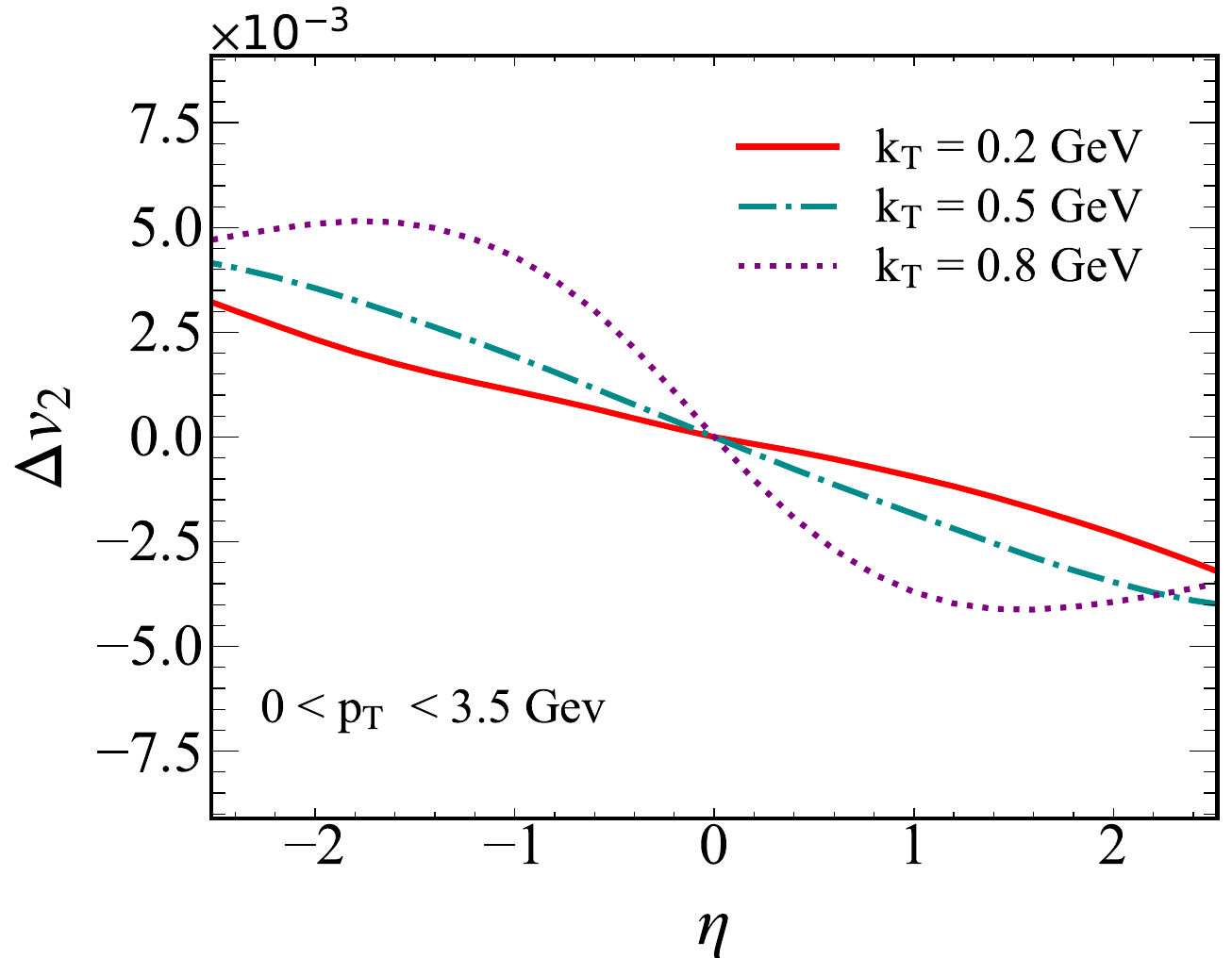}
\includegraphics[width=0.85\linewidth]{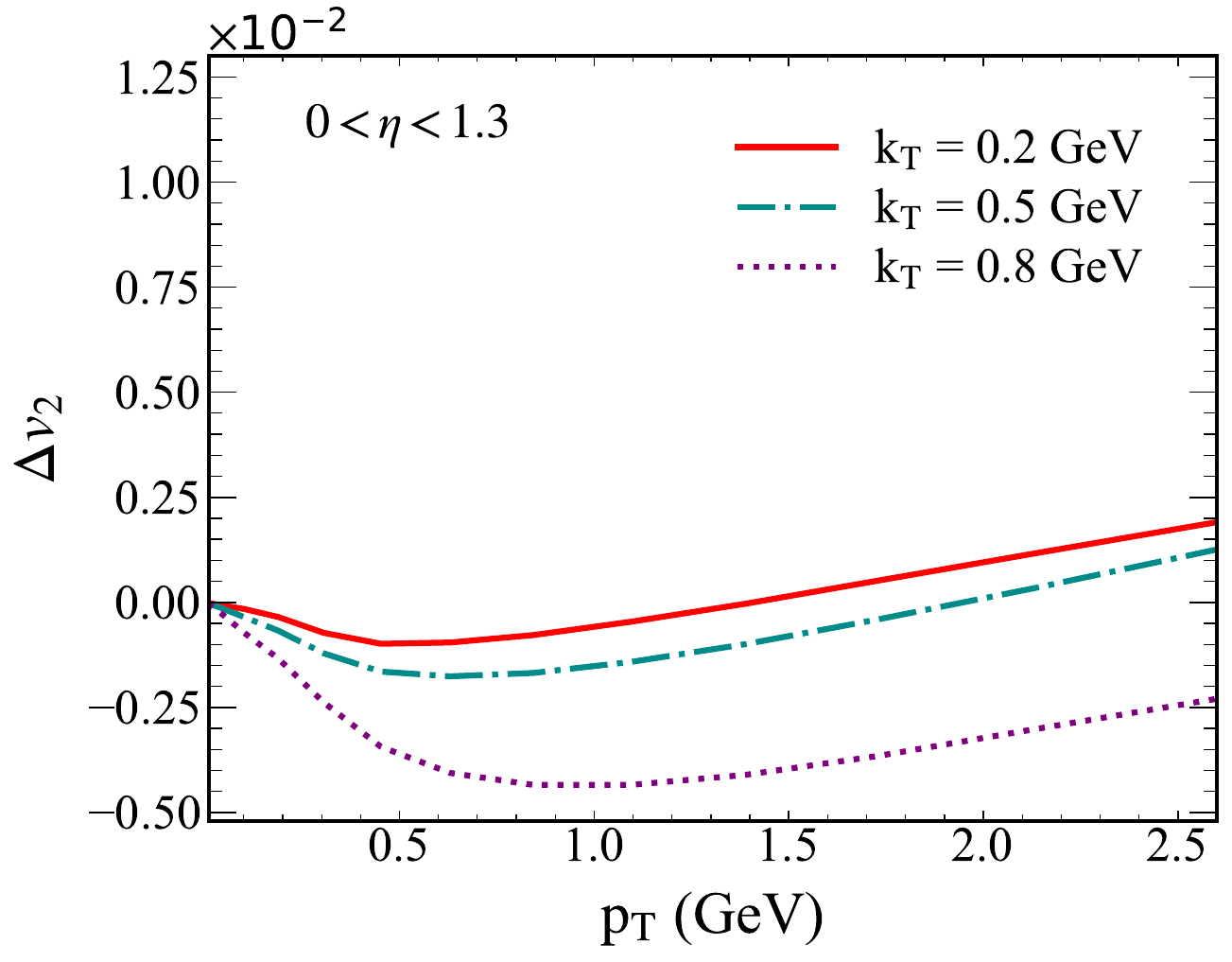}                         
\caption{(Color online) $\Delta v_{2}(\eta)$ (upper panel) and $\Delta v_{2}(\pt)$ (lower panel) distributions in Au+Au collisions at $\snn=200$~GeV (5-40\% centrality) for sub-nucleonic structure parameters $n_{c}=3,~10,~20$. }
\label{f:v1_kt}
\end{figure}

In Fig.~\ref{f:v1_kt}, we systematically explore the role of the parton transverse momentum scale ($k_\text{T}$), which governs the initial-state tilt geometry of the quark-gluon plasma (QGP) fireball, on $\Delta v_{2}$. The upper panel illustrates the $\eta$ dependence of $\Delta v_{2}$, where $\Delta v_{2}(\eta)$ exhibits sensitivity to $k_\text{T}$ for $|\eta| < 1$. Larger $k_\text{T}$ values enhance $\Delta v_{2}$, attributed to the increased tilt of the QGP fireball in the reaction plane, which amplifies $v_{1}$ and consequently $\Delta v_{2}$ (refer to Eq.~(\ref{eq21})). Conversely, the lower panel reveals an inverse relationship between $k_\text{T}$ and $\Delta v_{2}(\pt)$, as larger $k_\text{T}$ suppresses $\Delta v_{2}(\pt)$ due to a reduction in triangular flow ($v_{3}$). This $k_\text{T}$-dependent modulation arises from the altered initial-state geometry and subsequent hydrodynamic evolution of the QGP.

\begin{figure}[!t]
\centering
\includegraphics[width=0.85\linewidth]{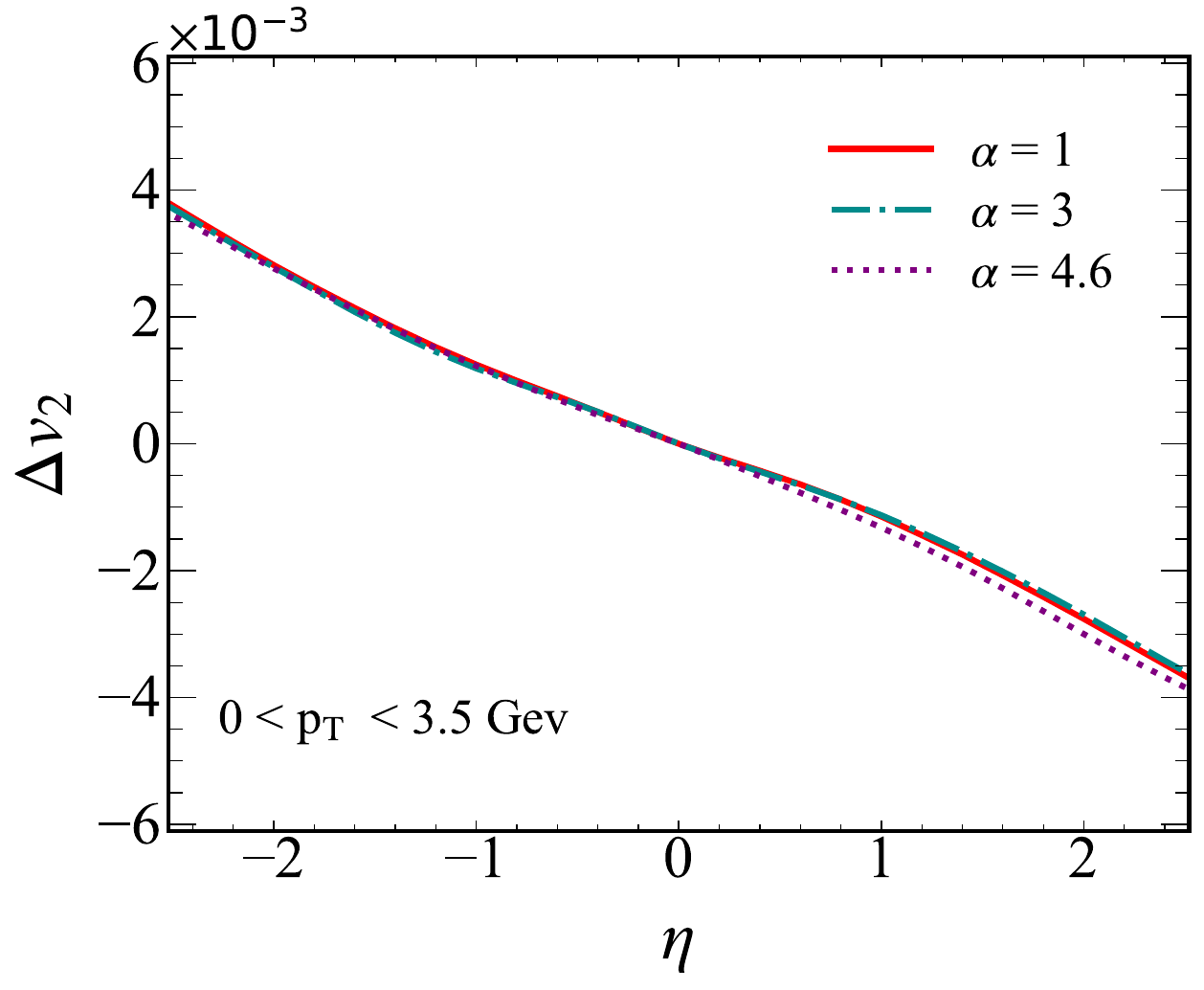}
\includegraphics[width=0.85\linewidth]{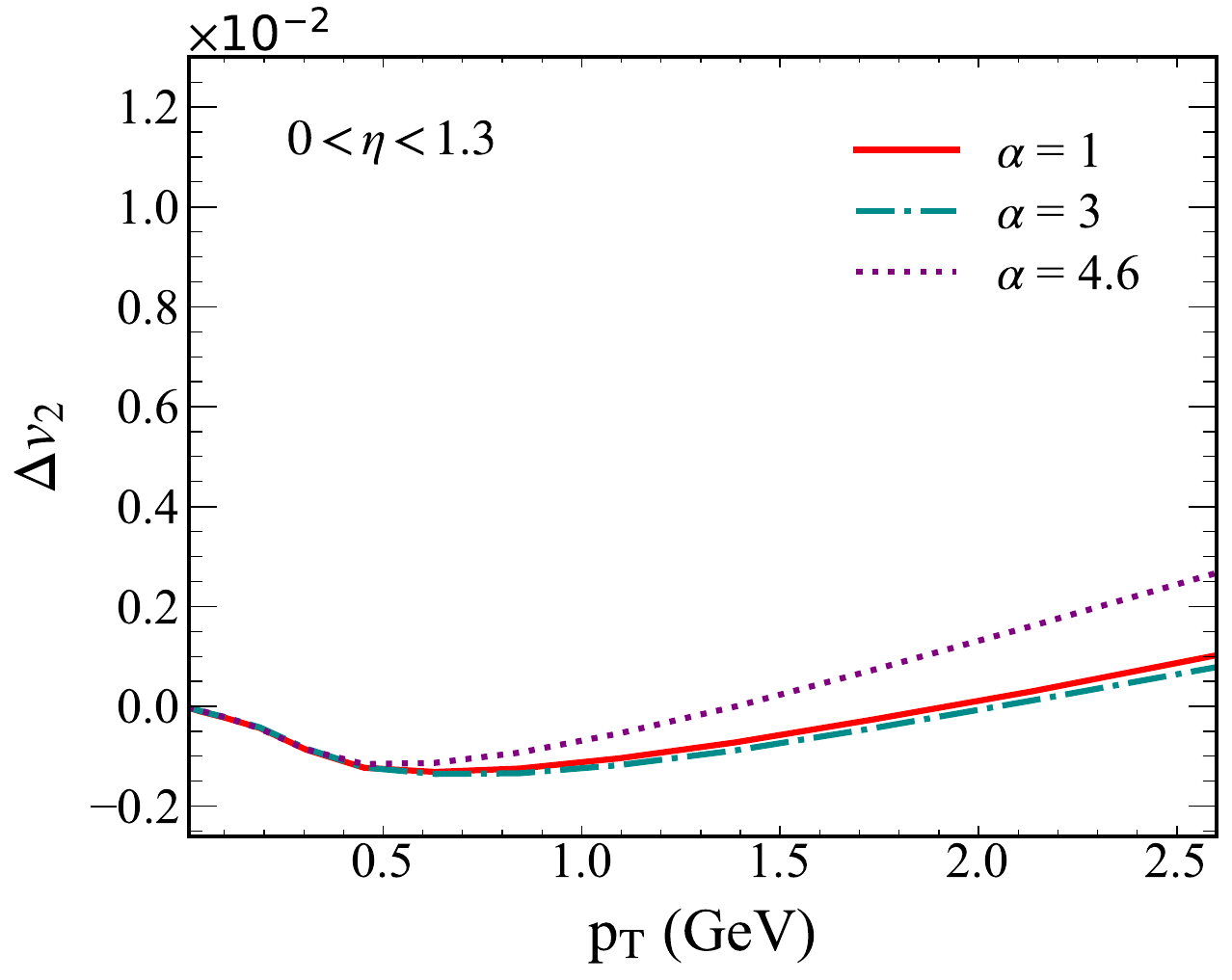}
\caption{(Color online) $\Delta v_{2}(\eta)$ (upper panel) and $\Delta v_{2}(\pt)$ (lower panel) distributions in Au+Au collisions at $\snn=200$~GeV (5-40\% centrality) for the shape of the fragmentation profile parameters $\alpha =1,~3,~4.6$.}
\label{f:v1_alpha}
\end{figure}

In Fig.~\ref{f:v1_alpha}, we investigate the impact of the fragmentation profile shape parameter $\alpha$ on $\Delta v_{2}$. The parameter $\alpha$ controls the longitudinal distribution of the QGP fireball in the forward/backward rapidity regions. As shown in the upper panel, $\Delta v_{2}(\eta)$ exhibits minimal sensitivity to $\alpha$ within $|\eta| < 2.5$, indicating that $\Delta v_{2}$ in the central rapidity region is governed by other dynamical factors. The lower panel reveals a non-monotonic dependence of $\Delta v_{2}(\pt)$ on $\alpha$: $\Delta v_{2}(\pt)$ initially decreases with $\alpha$, and then increases for larger $\alpha$. This behavior arises from the dual role of $\alpha$ in modulating triangular flow ($v_{3}$), where moderate $\alpha$ suppresses $v_{3}(\pt)$, while larger $\alpha$ enhances it.

\begin{figure}[!t]
\centering
\includegraphics[width=0.85\linewidth]{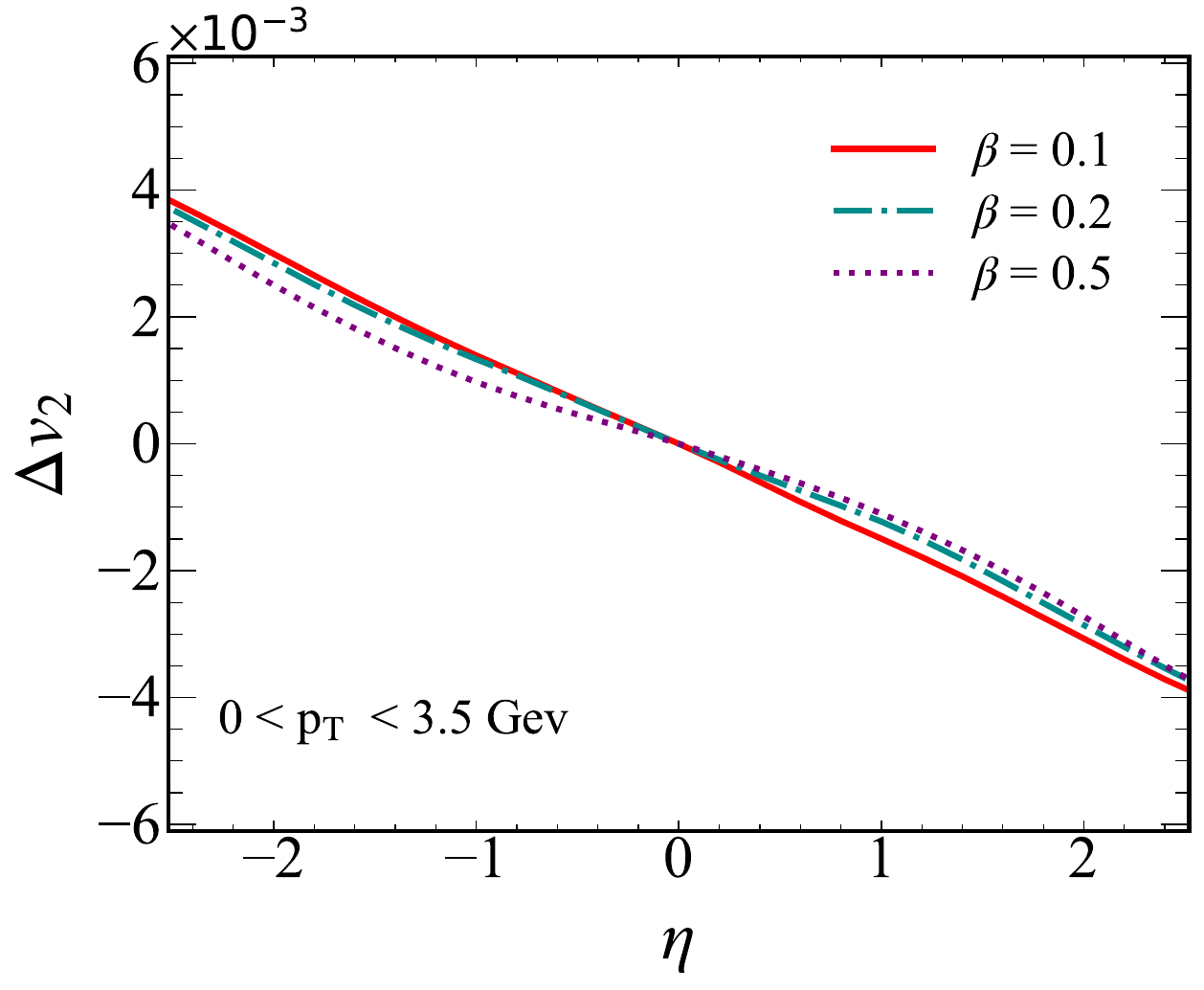}
\includegraphics[width=0.85\linewidth]{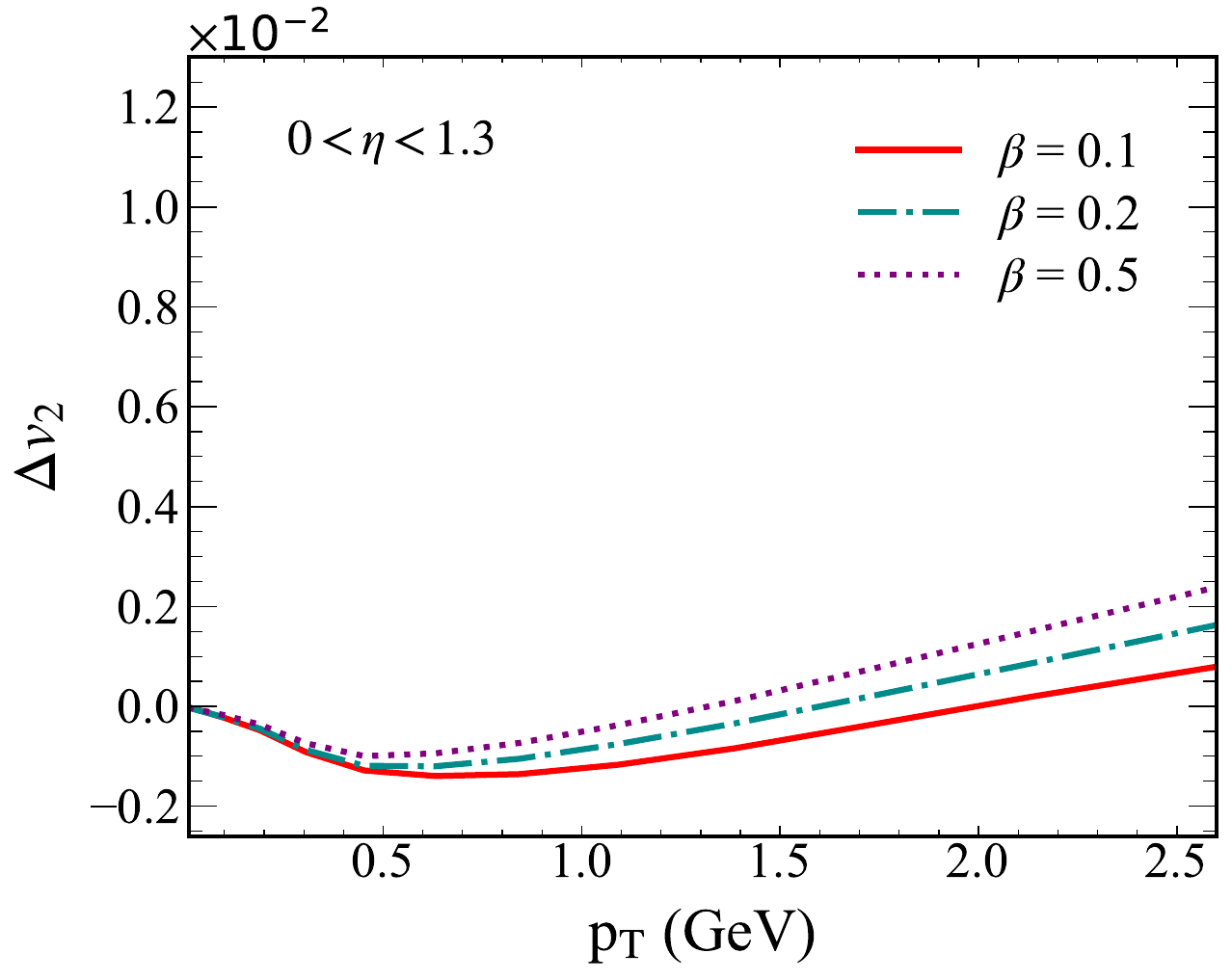}
\caption{(Color online) $\Delta v_{2}(\eta)$ (upper panel) and $\Delta v_{2}(\pt)$ (lower panel) distributions in Au+Au collisions at $\snn=200$~GeV (5-40\% centrality) for the shape of the fragmentation profile parameters $\beta =0.1,~0.2,~0.5$.}
\label{f:v1_beta}
\end{figure}

In Fig.~\ref{f:v1_beta}, we examine the effect of the fragmentation profile shape parameter $\beta$ on $\Delta v_{2}$. Similar to $\alpha$, $\beta$ governs the longitudinal geometry of the QGP fireball. As shown in the upper panel, $\Delta v_{2}(\eta)$ changes monotonically with increasing $\beta$ within $|\eta| < 2.5$, as larger $\beta$ values produce a flatter longitudinal distribution, reducing initial-state anisotropy. The lower panel reveals that samller $\beta$ values correspond to a reduced $\Delta v_{2}(\pt)$, as a flatter initial-state fragmentation profile diminishes anisotropic flow development.

From Figs.~\ref{f:v1_nc}--\ref{f:v1_beta}, we conclude that $\Delta v_{2}(\pt)$ is sensitive to $n_{c}$, $k_\text{T}$, $\alpha$, and $\beta$, making it as a valuable probe for exploring the initial state of the QGP and for refining and constraining parameters in the TRENTo-3D initial condition model.

\begin{figure}[!t]
\centering
\includegraphics[width=0.85\linewidth]{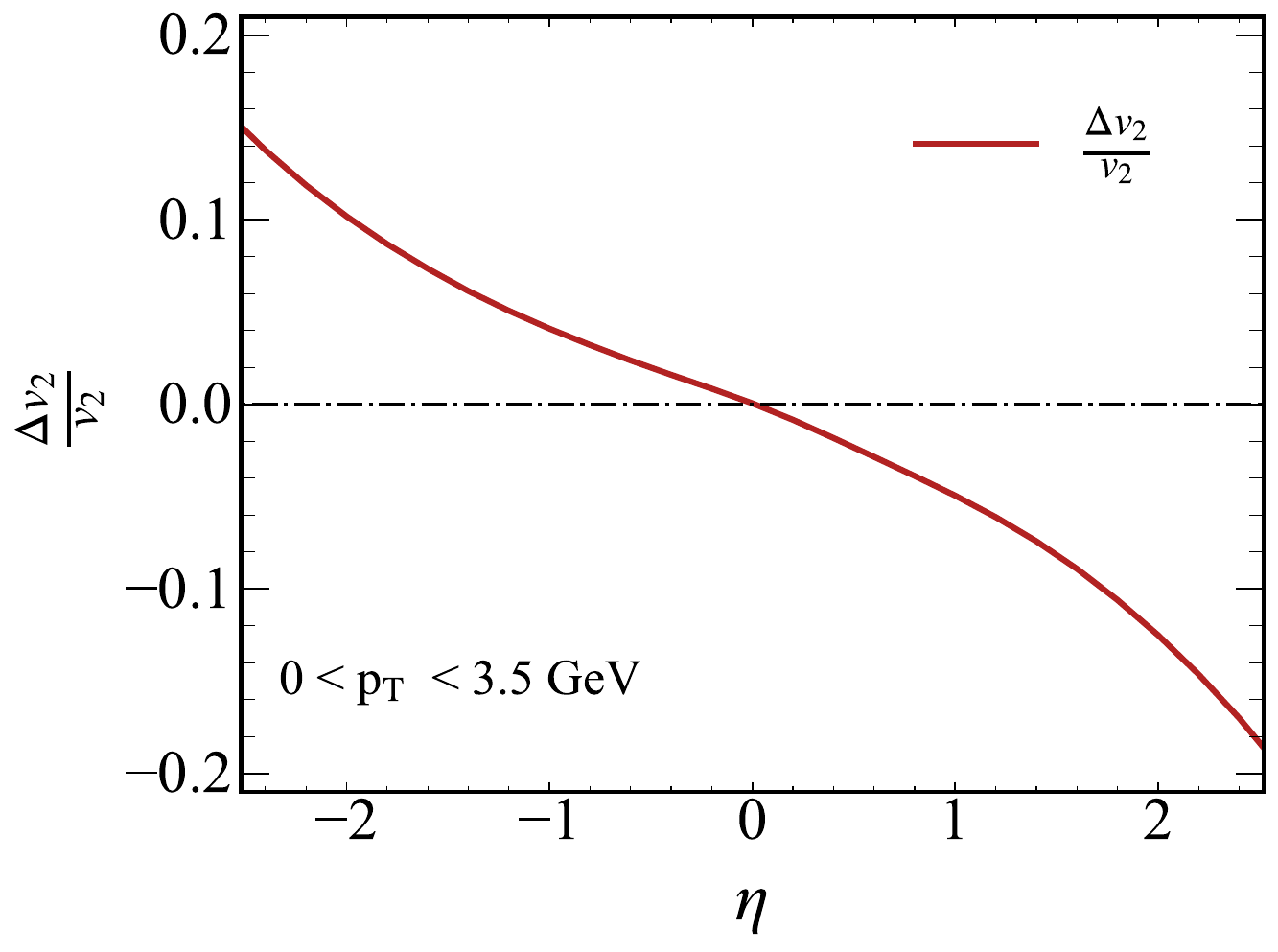}\\
\includegraphics[width=0.85\linewidth]{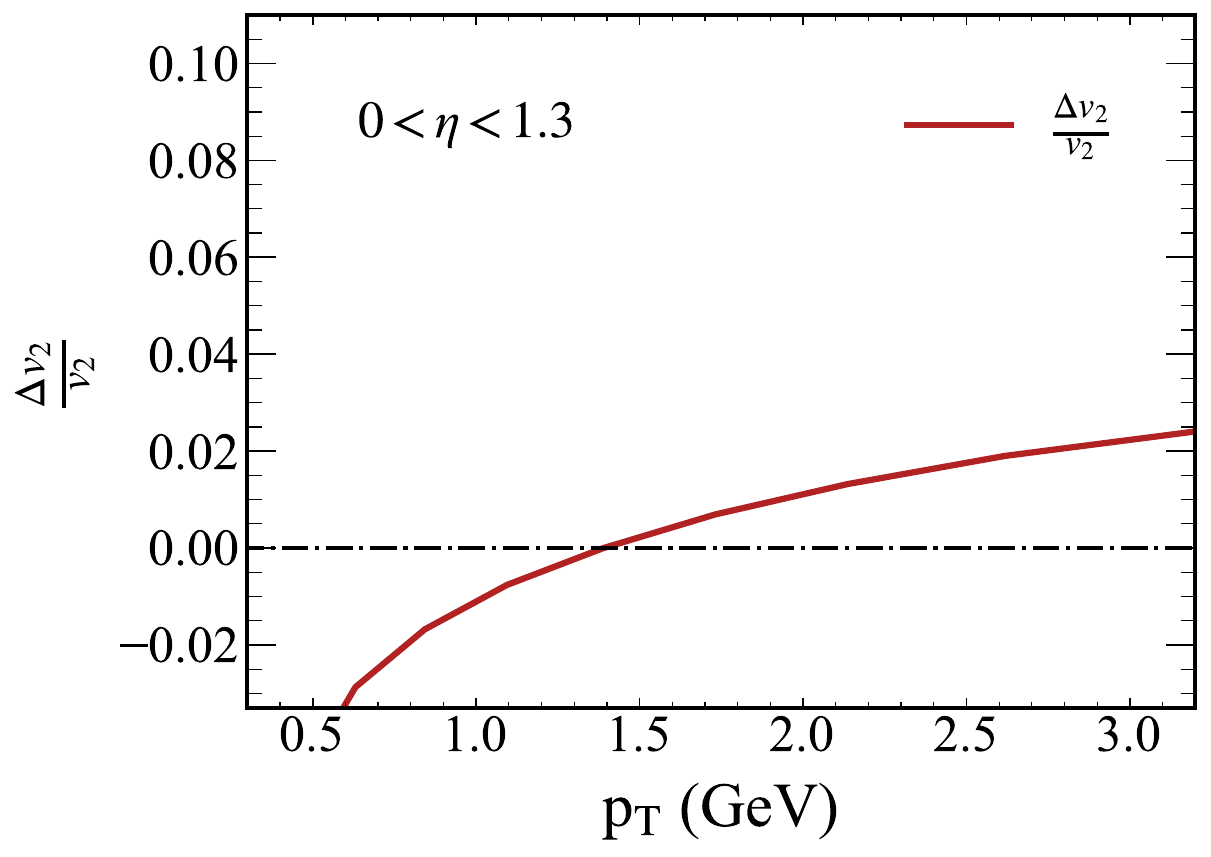}
\caption{(Color online) $\Delta v_2/v_2(\eta)$ (upper panel) and $\Delta v_2/v_2(\pt)$ (lower panel) distributions in Au+Au collisions at $\snn=200$~GeV (5-40\% centrality).}
\label{f:v1_v2ratio}
\end{figure}

It is well studied that various systematic uncertainties in the transverse initial condition models can lead to significant variations in the elliptic flow $v_{2}$~\cite{Shen:2011zc,Ruggieri:2013bda}. These model dependencies can also affect predictions of the elliptic flow difference $\Delta v_{2}$. Previous studies~\cite{Zhang:2021cjt,Parida:2022lmt} have demonstrated that scaling $\Delta v_{2}$ by $v_{2}$ can mitigate these systematic uncertainties, providing tighter constraints on the initial state of the quark-gluon plasma (QGP). 

In Fig.~\ref{f:v1_v2ratio}, we present predictions of $\Delta v_{2}/v_{2}$ as a function of pseudorapidity $\eta$ (upper panel) and transverse momentum $\pt$ (lower panel). We observe that the slope of $d[\Delta v_{2}/v_{2}]/d\eta$ reaches a value 4.4\% for $|\eta| < 1.0$, indicating a strong dependence of the scaled elliptic flow difference on pseudorapidity in the central rapidity region. Additionally, we find that $\Delta v_{2}/v_{2}$ cross the zero-point at $\pt \sim 1.5$ GeV, a result that can be tested against future experimental data. These findings suggest the potential of $\Delta v_{2}/v_{2}$ as a useful observable for probing the initial-state geometry and dynamics of heavy-ion collisions.

\section{Summary}
\label{v1section4}

Using the TRENTo-3D initial condition model coupled with (3+1)-dimensional CLVisc hydrodynamic simulations, we have revisited the left-right splitting of the elliptic flow ($\Delta v_{2}$) in Au+Au collisions at $\sqrt{s_{NN}} = 200$ GeV. We find that $\Delta v_{2}$ is primarily driven by odd flow harmonics, with the directed flow ($v_{1}$) being the dominant contributor to $\Delta v_{2}(\eta)$, consistent with previous transport and hydrodynamic model studies. However, unlike earlier findings, our results reveal that triangular flow ($v_{3}$) plays a significant role in $\Delta v_{2}(\pt)$, particularly in the rapidity region $|\eta| < 1.3$, where it becomes the dominant contributor.

To explore the sensitivity of $\Delta v_{2}$ to the initial state, we systematically investigated the influence of sub-nucleonic degrees of freedom ($n_c$, or ``hotspots'' number), transverse momentum scale ($k_\text{T}$), and fragmentation profile parameters ($\alpha$ and $\beta$). Our analysis shows that $\Delta v_{2}(\pt)$ is sensitive to all three effects in the TRENTo-3D initialization. In contrast, $\Delta v_{2}(\eta)$ exhibits strong sensitivity to $k_\text{T}$, which induces a tilted geometry of the QGP fireball in the initial state, but minimal sensitivity to $n_c$, $\alpha$, and $\beta$. These findings highlight $\Delta v_{2}$ as a valuable observable for probing the initial state of the QGP and constraining parameters in the TRENTo-3D model.

Finally, we present predictions for $\Delta v_{2}/v_{2}$ as a function of $\eta$ and $\pt$. We find that the slope $d[\Delta v_{2}/v_{2}]/d\eta$ reaches 4.4\% for $|\eta| < 1.0$, and $\Delta v_{2}/v_{2}$ cross zero-point at $\pt \sim 1.5$ GeV. Our study demonstrates that $\Delta v_{2}$ can complement $v_{1}$ and $v_{3}$ in constraining the three-dimensional profile of the QGP fireball created in heavy-ion collisions.

\begin{acknowledgements}
We thank Weiyao Ke for useful comments. This work was supported by the National Natural Science Foundation of China (NSFC) under Grant No.~12305138, the Guangdong Major Project of Basic and Applied Basic Research under Grant No. 2020B0301030008 and the Xiaogan Natural Science Foundation under Grant No.~XGKJ2023010063. Duan She's research is funded by the 2024 Henan Province International Science and Technology Cooperation Projects (No. 242102521068). 
\end{acknowledgements}

\bibliographystyle{unsrt}
\bibliography{clv3}

\end{document}